\documentstyle[preprint,aps]{revtex}

\newcommand{\be}{\begin{equation}}
\newcommand{\ee}{\end{equation}}
\newcommand{\bea}{\begin{eqnarray}}
\newcommand{\eea}{\end{eqnarray}}
\newcommand{\non}{\nonumber}

\newtheorem{prop}{Proposition}[section]
\newcommand{\ra}{\rangle}

\newcommand{\lam}{\lambda} 
\newcommand{\Lam}{\Lambda} 
 
\newcommand{\al}{\alpha}

\begin{document}
\tightenlines 
\title{Construction of some missing eigenvectors of the XYZ spin chain 
at the discrete coupling constants and 
the exponentially large spectral degeneracy of the transfer matrix}
\author{ Tetsuo Deguchi \footnotetext[1]{deguchi@phys.ocha.ac.jp .} }
%
\address{ 
Department of Physics, Faculty of Science \\
      Ochanomizu University \\
         2-1-1 Ohtsuka, Bunkyo-Ku, Tokyo 112-8610, Japan .}   
\preprint{OCHA-SP-01-07} 
\maketitle
\begin{abstract} 
We discuss an algebraic method for constructing eigenvectors of 
the transfer matrix of the eight-vertex model at the discrete coupling 
parameters.   We consider the algebraic Bethe ansatz of 
the elliptic quantum group $E_{\tau, \eta}(sl_2)$   
for the case where the parameter $\eta$ satisfies  
$2 N \eta = m_1 + m_2 \tau $ for arbitrary integers $N$, $m_1$ and $m_2$.
When  $m_1$ or  $m_2$ is odd, the eigenvectors thus obtained 
have not been discussed previously.   
Furthermore, we  construct a family of degenerate eigenvectors 
of the XYZ spin chain, some of which are shown to be related to  
the $sl_2$ loop algebra symmetry of the XXZ spin chain.  
We show that the dimension of some degenerate eigenspace  
of the XYZ spin chain on $L$ sites is given by 
 $N \, 2^{L/N}$, if $L/N$ is an even integer. 
The construction of eigenvectors of the transfer matrices 
of some related IRF (interaction-round-a-face) models is also discussed.  
\end{abstract}
%
%
 \newpage
  \setcounter{equation}{0} 
  \renewcommand{\theequation}{1.\arabic{equation}}
\section{Introduction}   
 The exact solution \cite{Baxter0,Baxter1,Baxter2,Baxter3} 
 of the eight-vertex model has played 
 a central role in the study 
 of exactly solved models and integrable lattice models 
 \cite{Baxter-book,Jimbo-book}. 
  The partition function of the model was 
  obtained  by the functional method 
  of the transfer matrix \cite{Baxter0}. 
  The eigenvectors of the transfer matrix of the model 
  was constructed  through the vertex-IRF correspondence 
  \cite{Baxter1,Baxter2,Baxter3}.  
 The algebraic Bethe ansatz was formulated   
 for the eight-vertex model \cite{8VABA}.  
 The knowledge of the exact solution 
  is also important in the continuum limits 
  of the lattice models to the field theories 
 and their connections to conformal field theories 
 \cite{Krinsky,Luther,Pokrovsky,Affleck}.  

\par 
 Recently,  another algebraic Bethe ansatz method has been introduced  
 for the eight-vertex model \cite{Felder,FV1,FV2}. 
 The method is  based on the elliptic quantum group $E_{\tau,\eta}(sl_2)$\cite{Felder,FV1}, 
  which is  associated with the $R$-matrix 
 of the eight-vertex solid-on-solid model (8VSOS model)  \cite{Baxter2,ABF},  
 and is also related to Drinfeld's quasi-Hopf algebra 
 \cite{BBB,Jimbo}. In Ref. \cite{FV2}, however, 
 the eigenvectors of the transfer matrix of the eight-vertex model 
 is discussed only for the case where the parameter $\eta$ is generic.  
Here we note that both for the generic and discrete $\eta$ cases 
some eigenvectors are discussed in Refs. \cite{Baxter1,Baxter2,Baxter3,8VABA}.    
 Thus, the  primary purpose of this paper is  to discuss 
 the algebraic Bethe ansatz of $E_{\tau,\eta}(sl_2)$ 
 at the discrete coupling constants. Then, we show that 
 it also gives such eigenvectors of the eight-vertex model 
 that are not discussed in Refs. \cite{Baxter1,Baxter2,Baxter3,8VABA}.

  \par Let us discuss the coupling parameters 
  $\tau$ and $\eta$ of $E_{\tau,\eta}(sl_2)$, 
  explicitly. For an illustration, we consider 
 the Hamiltonian of the  XYZ  spin chain 
 under the periodic boundary conditions, which 
 is given by the derivative of the homogeneous  
transfer matrix of the eight-vertex model  
 \be 
H_{XYZ} = \sum_{j=1}^{L} \left( J_X \sigma_j^X \sigma_{j+1}^X +
J_Y \sigma_j^Y \sigma_{j+1}^Y + J_Z \sigma_j^Z \sigma_{j+1}^Z  \right) 
\ee
In terms of the elliptic modulus $k$ (or $\tau$) and the coupling parameter $\eta$, 
the coupling constants of the XYZ chain 
 are expressed as follows
\be 
J_X  =  J(1 + k \, {\rm sn}^2( 2\eta,k) ) \, , \quad 
J_Y  =  J(1 - k \, {\rm sn}^2 ( 2\eta,k) ) \, , \quad 
J_Z  =  J \, {\rm cn}(2\eta,k) {\rm dn}(2\eta,k) 
\label{JXYZ}
\ee 
Here ${\rm sn}(z,k)$, ${\rm cn}(z,k)$ and  ${\rm dn}(z,k)$
 denote the Jacobian elliptic functions. 
For  the discrete cases of Refs. \cite{Baxter1,Baxter2,Baxter3,8VABA},   
 the eigenvectors of the transfer matrix of the eight-vertex model 
 are constructed  under the condition: 
$ 2N \eta = 4 m_1 K + 2 i m_2 K^{'}$ where $N$, $m_1$ and $m_2$ are 
arbitrary integers. 
Here, the symbols $K$ and $K^{'}$ have denoted the 
complete elliptic integrals of the first and second kind, 
respectively.  It has not been discussed explicitly  
how one can construct eigenvectors for the case 
of $2N \eta = 2 m_1 K + i m_2 K^{'}$ 
with given integers $N$, $m_1$ and $m_2$. 
Here we note that  the functional relation \cite{Baxter0}  
 is derived under  the latter condition: 
 $2N \eta = 2 m_1 K + i m_2 K^{'}$.

\par 
There is another motivation of the paper. 
Recently, it has been found  that the 
XXZ spin chain  at the  roots of unity 
 has the spectral symmetry 
 of the $sl_2$ loop algebra \cite{DFM}. 
 The explicit expressions of the generators commuting with 
 the XXZ Hamiltonian under the periodic boundary conditions 
are given in  Appendix A. 
Several non-trivial properties of the spectral degeneracy 
of the $sl_2$ loop algebra 
have been discussed \cite{FM}.  Here, 
the Hamiltonian of the XXZ spin chain gives   
the special case of the XYZ Hamiltonian with $J_X = J_Y$, 
 which is obtained  by taking the trigonometric limit:
  $k \rightarrow 0$ for the 
 coupling constants of eq. (\ref{JXYZ}). Furthermore, 
it has been suggested in Ref. \cite{DFM} 
through a numerical study that 
the XYZ spin chain should have a very large  spectral degeneracy 
similar to the $sl_2$ loop algebra symmetry of the XXZ spin chain. In fact, 
the discrete condition:  $2N \eta = 2 m_1 K + i m_2 K^{'}$ of the XYZ spin chain 
corresponds to the roots of unity condition of the XXZ Hamiltonian,  
in the trigonometric limit.  
Therefore, the  construction of eigenvectors 
of the XYZ spin chain at the discrete coupling constants 
 should be important in studying 
  the conjectured spectral degeneracy of the XYZ spin chain.

\par 
The outline of the paper is given in the following. 
In \S 2 we introduce the elliptic quantum group 
 $E_{\tau,\eta}(sl_2)$, briefly. In the paper 
 we employ almost the same symbols with Ref. \cite{FV2} 
 for $E_{\tau, \eta}(sl_2)$.      
In \S 3 we show the main results of   
the algebraic Bethe ansatz of $E_{\tau,\eta}(sl_2)$ 
 for the discrete $\eta$ case of $2N\eta = m_1 + m_2 \tau$, 
which corresponds to the case of  $2N \eta = 2 m_1 K + i m_2 K^{'}$  
in the notation of Refs. \cite{Baxter0,Baxter1,Baxter2,Baxter3}.  
Then, we obtain a slightly generalized expression for 
the  eigenvalues of the transfer matrix  expressed 
in terms of rapidities.  In \S 4, 
 for the discrete $\eta$ case, we discuss some important points in   
 constructing  eigenvectors for the 8VSOS model and 
 its two variants: the  ABF (or 8VRSOS) model and the cyclic SOS model 
 \cite{Pearce,Kuniba,Akutsu}, 
 and then for the eight-vertex model. 
In \S 5, we formulate an algebraic  
method for constructing degenerate eigenvectors 
of the transfer matrix of the eight-vertex model. 
It indeed gives a large number of 
degenerate eigenvectors.  For an illustration, 
we calculate analytically the dimension of the largest degenerate eigenspace 
of the XYZ model defined on the $L$ sites. We show that it is given by   
  $N \, 2^{L/N}$,  if $L/N$ is an even integer. 
Thus we see that the spectral degeneracy can increase 
exponentially with respect to the lattice size $L$.  
Finally, in \S 6, we give some discussions. 
In order to make the paper self-consistent, 
some appendices are provided. In Appendix A, 
we review the $sl_2$ loop algebra 
symmetry of the XXZ spin chain at the root of unity. 
In Appendix B, we explain the evaluation modules 
of the elliptic quantum group 
$E_{\tau, \eta}(sl_2)$. Finally in Appendix C, we give the 
Boltzmann weights of  the RSOS models 
associated with the eight-vertex model.

  \setcounter{equation}{0} 
  \renewcommand{\theequation}{2.\arabic{equation}}
\section{Elliptic quantum group $E_{\tau,\eta}(sl_2)$}

\par 
The elliptic quantum group $E_{\tau,\eta}(sl_2)$ 
 is an algebra generated by meromorphic functions of a variable 
 $h$ and the matrix elements of an operator-valued  matrix $L(z, \lam)$ 
 with non-commutative entries \cite{Felder,FV1,FV2}, 
 which satisfy the Yang-Baxter relation with a dynamical shift 
\bea 
& & R^{(12)}(z_{12}, \lam - 2\eta h^{(3)}) L^{(1)}(z_1,\lam ) 
L^{(2)}(z_2,\lam - 2\eta h^{(1)}) \non \\
&=&  L^{(2)}(z_2, \lam) 
L^{(1)}(z_1, \lam- 2 \eta h^{(2)}) R^{(12)}(z_{12}, \lam)
\label{DYBR}
\eea
Here $h$ is a  generator of  the Cartan subalgebra ${\bf h}$ 
of $sl_2$.

\par 
Let us formulate the $R$-matrix of the elliptic quantum group $E_{\eta, \tau}(sl_2)$, 
explicitly.   We introduce the theta function 
\be 
\theta(z; \tau) = 2 p^{1/4} \sin \pi z \prod_{n=1}^{\infty}
(1-p^{2n})(1-p^{2n}\exp(2\pi i z)) (1-p^{2n}\exp(-2\pi i z)) \, , 
\ee
where the nome $p$ is related to  the parameter $\tau$ by $p=\exp(\pi i \tau)$  
with  ${\rm Im} \quad \tau>0$ .
Let $V$ be the two-dimensional complex vector space with 
the basis $e[1]$ and $e[-1]$. Here we denote $e[-1]$ also   
 as $e[2]$,  
and let $E_{ij}$ denote the matrix satisfying 
 $E_{ij} e[k] = \delta_{jk} e[i]$.  
Then, the $R$-matrix $R(z, \lam) \in End(V)$ is given by 
\bea 
R(z,\lam; \eta, \tau) & = & E_{11} \otimes E_{11} + E_{22} \otimes E_{22} 
 + \alpha(z, \lam) E_{11} \otimes E_{22} \non \\
&&  + \beta(z, \lam) E_{12} \otimes E_{21}  + \beta(z, -\lam) E_{21} \otimes E_{12}
    + \alpha(z, -\lam) E_{22} \otimes E_{11}
\label{R-matrix}
\eea
where $h=E_{11} - E_{22}$ and  
$\alpha(z, \lam)$ and $\beta(z, \lam)$ are defined by 
\be 
\alpha(z, \lam) = {\frac {\theta(z)\theta(\lam+2\eta)} {\theta(z-2\eta)\theta(\lam)}} \, , 
\qquad \beta(z, \lam) = - 
{\frac {\theta(z+\lam)\theta(2\eta)} {\theta(z-2\eta)\theta(\lam)}} \, . 
\ee

\par 
The elliptic quantum group $E_{\tau, \eta}(sl_2)$ has  
 two types of generators: (i) 
meromorphic functions of $f(h)$ of one-variable 
with period $1/\eta$; (ii) the non-commutative matrix elements 
of the $L$ operator: 
$a(z, \lam)$, $b(z, \lam)$, $c(z, \lam)$, and  $d(z, \lam)$, 
which satisfy the relations 
arising from (\ref{DYBR}), such as shown in the following 
\bea 
b(z_1, \lam)b(z_2, \lam +2\eta) & = & b(z_2, \lam)b(z_1, \lam +2\eta) \non \\    
b(z_1, \lam)a(z_2,\lam+2\eta) & = & b(z_2,\lam)a(z_1, \lam+2\eta) 
\beta(z_1-z_2, \lam) 
+ a(z_2, \lam)b(z_1, \lam-2\eta) \alpha(z_1-z_2, -\lam) \non \\ 
\label{fcr}
\eea     
Hereafter, we suppress the $\lam$-dependence 
of the operators $a(z, \lam)$, $b(z, \lam)$, $c(z, \lam)$ 
and  $d(z, \lam)$, for  simplicity.  
 
 \par 
 The definition of the evaluation Verma module for the elliptic quantum group 
 $E_{\tau, \eta}(sl_2)$ is explicitly given in Appendix B.

   \setcounter{equation}{0} 
  \renewcommand{\theequation}{3.\arabic{equation}}
\section{Algebraic Bethe ansatz at the discrete coupling parameters}   
 
\par 
Let $W=V_{\Lam_1}(z_1) \otimes \cdots \otimes V_{\Lam_n}(z_n)$ 
be the tensor product of evaluation modules $V_{\Lam_j}(z_j)$'s \cite{FV2}
(See also Appendix B). The total spin $\Lam$ is given by     
$\Lam=\Lam_1 + \Lam_2 + \cdots + \Lam_n$.  Here, we assume $\Lam$ 
is given by  $\Lam = 2\ell$ or $\Lam = 2\ell+1 $ with an integer $\ell$.  
The transfer matrix  $T(z)$ of the elliptic algebra 
is given by the trace of the $L$-operator acting on the tensor product space 
$W$.   We denote by $v_0$  the highest weight vector of $W$: $h v_0 = \Lam \, v_0$. 

\par 
Let us assume hereafter  that 
the coupling parameter $\eta$ is given by  the discrete values:
$2 N \eta = m_1 + m_2 \tau$ 
for some integers $N$, $m_1$ and $m_2$. 
We now take an integer $m$ satisfying the following condition
\be 
2m = \Lam - r N \, , \qquad {\rm for } \quad r \in {\bf Z} \, . 
\label{rN} 
\ee 
We consider the $m$th product of the creation operators:
$b(t_1) \cdots b(t_m)$ with $m$ rapidities $t_1, \ldots, t_m$. 
Here we note that the integers $m$ and $\ell$ can be different,  
while in Ref. \cite{FV2} only the case of $m=\ell$ and generic $\eta$ 
is discussed.  Let us define a vector $v_c$ 
by $v_c=g_c(\lam) v_0$ where $g_c(\lam)$ is given by 
\be 
g_c(\lam) = e^{c\lam} \, 
\prod_{j=1}^{m} {\frac {\theta(\lam-2 \eta j)}{\theta(2 \eta)}} 
\ee

\par 
Let us first consider the case when $m_2=0$. 
Then, through the fundamental commutation relations such as shown in eq. (\ref{fcr}), 
we can show 
\be
T(w) \, b(t_1) \cdots b(t_m) v_c 
= C_0(w) \, b(t_1) \cdots b(t_m) v_c 
+ \sum_{j=1}^{m} C_j \, b(t_1) \cdots b(t_{j-1}) b(w) b(t_{j+1}) \cdots b(t_m) v_c \, . 
\label{Tb}
\ee
Here the coefficients $C_0$ and $C_j$ are given by 
\bea 
C_0(w) & = & e^{-2 \eta c} \, \prod_{j=1}^{m} 
{\frac {\theta(w-t_j+2 \eta)}  {\theta(w-t_j)}}
+ (-1)^{r m_1} e^{2 \eta c} \, \prod_{j=1}^{m} 
{\frac {\theta(w-t_j -2 \eta)}  {\theta(w-t_j)}}
\prod_{\al=1}^{n} 
{\frac {\theta(w- p_{\al})}  {\theta(w- q_{\al})}}
\label{eigenvalue} 
\non \\
C_j & = &  e^{-2\eta c} 
 {\frac {\theta(2 \eta )} {\theta(t_j -w)}} 
{\frac {\theta(t_j-w +\lam )} {\theta(\lam)}} 
\left(  \prod_{k=1;k \ne j}^{m}  f_{kj} - 
(-1)^{r m_1}   e^{4\eta c}\prod_{k=1;k \ne j}^{m}  f_{jk} \, \prod_{\al=1}^{n} 
{\frac {\theta(t_j-p_{\al})} {\theta(t_j-q_{\al})}}  
\right)
\eea
where $p_{\al} = z_{\al} + \eta (-\Lam_{\al} +1)$ and 
$q_{\al}= z_{\al} + \eta (\Lam_{\al} +1)$ for $\al=1, \ldots n$, 
 and the symbol $f_{jk}$ denotes  the following  
\be 
f_{jk}= {\frac {\theta(t_j -t_k - 2 \eta)} {\theta(t_j- t_k)}}
\ee
Thus, we have found that 
$b(t_1)b(t_2) \cdots b(t_m) v$ is an eigenvector of the transfer matrix 
$T(z)$ with the eigenvalue $C_0(z)$, if rapidities $t_1, t_2, \ldots, t_m$ 
satisfy the Bethe ansatz equations  
\be 
\prod_{k=1}^{n} {\frac {\theta(t_j-p_k)} {\theta(t_j-q_k)}} 
= (-1)^{rm_1} e^{-4\eta c} \prod_{k=1; k \ne j}^{m} 
{\frac {\theta(t_j-t_k + 2\eta)} {\theta(t_j-t_k -2 \eta)}} 
\qquad {\rm for} \quad j=1, \ldots, m \, .
\label{BAE} 
\ee
We remark that the factor $(-1)^{r m_1}$ can be  important 
for the case when $m_1$ is odd, which has not been 
discussed in the references \cite{Baxter1,Baxter2,Baxter3,8VABA}

\par 
When $m_2 \ne 0$, we consider 
the ``renormalization''of the  theta function  
\cite{Baxter1}
\be 
{\tilde \theta}(z) = \theta(z) \exp \left( \pi m_2(z-1/2)^2/(2N\eta) \right)  
\label{renor} 
\ee
Then we have 
\be 
{\tilde \theta}(z+2N \eta) = (-1)^{ m_1 (m_2+1) }  {\tilde \theta}(z) 
\ee
Replacing all the theta functions $\theta(z)$'s in the above discussions 
with the renormalized ones ${\tilde \theta}(z)$'s, 
 for the case of $2 N \eta = m_1 + m_2 \tau$,  we have  
\bea
T(w) \, b(t_1) \cdots b(t_m) v_c 
& = & {\tilde C}_0(w) \, b(t_1) \cdots b(t_m) v_c  \non \\
& & + \sum_{j=1}^{m} {\tilde C}_j \, b(t_1) \cdots 
b(t_{j-1}) b(w) b(t_{j+1}) \cdots b(t_m) v_c \, . 
\label{Tb2}
\eea
Here the coefficients ${\tilde C}_0$ and ${\tilde C}_j$ are given by the following 
\bea 
{\tilde C}_0(w) & = & e^{-2 \eta c} \, \prod_{j=1}^{m} 
{\frac { {\tilde \theta}(w-t_j+2 \eta)}  { {\tilde \theta}(w-t_j)}}
+ (-1)^{r m_1 (m_2+1)} e^{2 \eta c} \, \prod_{j=1}^{m} 
{\frac { {\tilde \theta}(w-t_j -2 \eta)}  { {\tilde \theta}(w-t_j)}}
\prod_{\alpha=1}^{n} 
{\frac { {\tilde \theta}(w- p_{\al})}  { {\tilde \theta}(w- q_{\al})}}
\non \\
{\tilde C}_j & = &  e^{-2\eta c} 
 {\frac { {\tilde \theta}(2 \eta )} { {\tilde \theta}(t_j -w)}} 
{\frac { {\tilde \theta}(t_j-w +\lam )} { {\tilde \theta}(\lam)}} \non \\ 
&& \times \, 
\left(  \prod_{k=1;k \ne j}^{m}  {\tilde f}_{kj} - 
(-1)^{r m_1 (m_2+1)}  e^{ 4 \eta c} \prod_{k=1;k \ne j}^{m}  {\tilde f}_{jk} \, \prod_{\al=1}^{n} 
{\frac { {\tilde \theta}(t_j-p_{\al})} { {\tilde \theta}(t_j-q_{\al})}}  
\right)
\eea
where  
${\tilde f}_{jk}= { {\tilde \theta}(t_j -t_k - 2 \eta)}/ { {\tilde \theta}(t_j- t_k)}$.  
 Thus, we have shown that 
$b(t_1) \cdots b(t_m) v$ is an eigenvector of the transfer matrix 
$T(z)$ with the eigenvalue ${\tilde C}_0(z)$, if rapidities $t_1, \ldots, t_m$ 
satisfy the Bethe ansatz equations  
\be 
\prod_{k=1}^{n} {\frac { {\tilde \theta}(t_j-p_k)} { {\tilde \theta}(t_j-q_k)}} 
= (-1)^{r m_1 (m_2+1)} e^{-4 \eta c} \prod_{k=1; k \ne j}^{m} 
{\frac { {\tilde \theta}(t_j-t_k + 2\eta)} { {\tilde \theta}(t_j-t_k -2 \eta)}} 
\qquad {\rm for} \quad j=1, \ldots, m \, .
\label{BAE2}
\ee
We remark  that the factor $(-1)^{r m_1 (m_2+1) }$ should be important 
when $m_1$ is odd and $m_2$ is even.

\par 
Hereafter in the paper, we discuss only the case of $m_2 = 0$ 
($2N\eta = m_1$) to each of the topics, for simplicity.  
However, the case of $m_2 \ne 0$ can be  discussed similarly 
by using the renormalized theta function (\ref{renor}).

  \setcounter{equation}{0} 
  \renewcommand{\theequation}{4.\arabic{equation}}
\section{The eight-vertex model and  related IRF models} 
 \subsection{The 8VSOS and 8VRSOS models}

\par 
Let us discuss  for the discrete $\eta$ case of $E_{\tau, \eta}(sl_2)$ 
how one can construct  eigenvectors  
of the transfer matrices of the 8VSOS model \cite{Baxter2}
and its restricted versions, the ABF (or 8VRSOS) model \cite{ABF} 
and the cyclic SOS model (8VCSOS model) \cite{Pearce,Kuniba,Akutsu}, 
 briefly.  The  RSOS models associated with the eight-vertex model 
 are reviewed in Appendix C.  

\par 
Let us assume that $\Lam_1 = \cdots = \Lam_n=1$, 
 hereafter in the paper. Thus, we have $n=\Lam$ 
 which is  given by $2 \ell $ or $2\ell +1$ 
 for an integer $\ell$. We also have $n$=$L$,  where  
  $L$ is the lattice size defined  in \S 1.  We shall denote $n$ and $\Lam$ by $L$, 
  hereafter. 

\par 
We introduce  the $L$-operator \cite{FV2} for the tensor product $W=V^{\otimes L}$  
 \be 
 L(z, \lam)= R^{(01)}(z-z_1, \lam-2\eta \sum_{j=2}^{L} h^{(j)}) 
   R^{(02)}(z-z_2, \lam - 2\eta \sum_{j=3}^{L} h^{(j)}) 
   \cdots  R^{(0L)}(z-z_L, \lam)  
\label{L-SOS}
 \ee
Here we recall that the transfer matrix $T(z)$ 
of $E_{\tau,\eta}(sl_2)$ for $W$ is given by 
the trace of the $L$-operator: $T(z, \lam)={\rm tr}_0 L(z, \lam)$. 

\par 
The $R$-matrix $R(z, \lam)$ of $E_{\eta, \tau}(sl_2)$ (which has been given in eq. (\ref{R-matrix}) )  
is related to the Boltzmann weights  $w(a,b,c,d; z)$  of the 8VSOS model  \cite{Baxter2} 
through the following relation  \cite{FV2} 
\be
R(z, -2\eta d) e[c-d] \otimes e[b-c] = \sum_{a} w(a,b,c,d; z) e[b-a] \otimes e[a-d] 
\label{R-w}
\ee   
Here $a,b,c,d$ denote the spin variables of the IRF (the Interaction Round a Face) model,  
which take integer values \cite{Baxter2}. The spin variables have  the constraint that 
the difference between the values of two nearest-neighboring spins should be  
given by  $\pm 1$.    Through the relation (\ref{R-w}), we can show that 
the transfer matrix $T(z)$ of $E_{\tau, \eta}(sl_2)$  
acting on the ``path basis'' corresponds 
to that of the 8VSOS model \cite{FV2}.  
Here we note that a ``path'' is given by  
a sequence of the values of spin variables 
satisfying the  constraints on adjacent spins.  
Thus, if we express the eigenvector $b(t_1)\cdots b(t_m) v_c$ 
of $E_{\tau,\eta}(sl_2)$ 
in terms of the path basis \cite{Baxter2,FV2}, then 
it gives that of the transfer matrix of the 8VSOS model.  

\par 
For the 8VRSOS model, the values of the spin variables 
 are restricted into $N$ values such as $0, 1, \ldots$, and $N-1$ \cite{ABF}.  
In the model,  it is not allowed for any pair of neighboring spins 
to have the values  0 and  $N-1$, respectively. 
Thus, only such paths of the 8VSOS model that satisfy the  conditions 
are allowed as paths of the 8VRSOS model.  
The  transfer matrix of the 8VRSOS model is defined on the restricted space of paths. 
We can construct eigenvectors of 8VRSOS model in the restricted  space of paths.

\par 
For the 8VSOS and ABF models,  the periodic boundary conditions 
on the path basis are satisfied  only when $L = 2m$. 
Thus, we can construct  the eigenvectors of the models 
only for the case when  $L$ is even. 

\par 
 For the cyclic SOS (or 8VCSOS) model \cite{Pearce,Kuniba,Akutsu}, 
 the spin variables 
take the restricted values such as $0, 1, \ldots, N-1$, where 
the values  0 and  $N-1$ are allowed  for any pair of neighboring spins.   
Thus, the admissibility condition on the spin values 
can be expressed by a cyclic graph of $N$ nodes. 
 The periodic boundary conditions on the path basis of the cyclic SOS model 
 are satisfied under the constraint: $L -2m=rN $, 
 where $r$ can be non-zero integers.  Therefore,     
 for the cyclic SOS model, we can consider eigenvectors also for 
the odd lattice-size case. For instance, we may take 
$L=7$ and consider the case of $N=3$, $r=1$ and $m=2$, 
which satisfies the constraint: $L-2m=rN$.  Furthermore, 
when $2N \eta= m_1$ with $m_1$ odd, then the 
factor $(-1)^{r m_1}$ in the formula of eigenvalues (\ref{eigenvalue}) 
is given by -1, not by 1. 
%
%

\subsection{The eight-vertex model}

Let us introduce the theta functions $\theta_0(z)$ and $\theta_1(z)$ satisfying 
$\theta_{\al}(z+1)= (-1)^{\al} \theta_{\al}(z)$ and 
 $\theta_{\al}(z+\tau)= i e^{-\pi i (z+ \tau/2)} \theta_{1-\al}(z)$ for $\alpha= 0,1$ , 
 where we define $\theta_1(z)$ by $\theta_1(z, \tau)= \theta(z, 2 \tau)$ \cite{FV2}.  
In terms of the theta functions,  
the $R$-matrix of the eight-vertex model \cite{Baxter0} is given by  
\bea 
R_{8V}(z)& = & a_{8V}(z) \left( E_{11} \otimes E_{11} + E_{22} \otimes E_{22} \right)
+ b_{8V}(z) \left( E_{11} \otimes E_{22} + E_{22} \otimes E_{11} \right) \non \\
& & + c_{8V}(z) \left( E_{12} \otimes E_{21} + E_{21} \otimes E_{12} \right)
+ d_{8V}(z) \left( E_{12} \otimes E_{12} + E_{21} \otimes E_{21} \right)
\eea
where the Boltzmann weights $a_{8V}(z), b_{8V}(z), c_{8V}(z)$ and $d_{8V}(z)$ are expressed as   
\bea 
a_{8V}(z)& = & {\frac {\theta_0(z) \theta_0(2\eta)} 
                {\theta_0(z- 2\eta) \theta_0(0)}} \, , \quad 
b_{8V}(z) = {\frac {\theta_1(z) \theta_0(2\eta)} 
                {\theta_1(z- 2\eta) \theta_0(0)}} \, ,  \non \\
c_{8V}(z) & = & - {\frac {\theta_0(z) \theta_1(2\eta)} 
                {\theta_1(z- 2\eta) \theta_0(0)}} \,, \quad 
d_{8V}(z) = - {\frac {\theta_1(z) \theta_1(2\eta)} 
                {\theta_0(z- 2\eta) \theta_0(0)}}  \, . 
\eea
The transfer matrix $T_{8V}(z)$ of the eight-vertex model is given by the trace 
over the 0th vector space: 
$T_{8V}(z) = tr_0 L_{8V}(z)$, where the operator $L_{8V}(z)$ is defined  by 
\be
L_{8V}(z) =  R_{8V}(z-z_1)^{(01)} \cdots   R_{8V}(z-z_L)^{(0L)}
\label{L-8V}
\ee

\par 
Let us consider  the correspondence between the eight-vertex model 
and the 8VSOS model \cite{Baxter2}. We introduce 
some symbols of Ref.  \cite{FV2} for $E_{\tau, \eta}(sl_2)$. 
We  define the following matrix  
\be 
S(z, \lam) = 
\left( 
\begin{array}{cc}
\theta_0(z-\lam+1/2) & - \theta_0(-z-\lam +1/2) \\  
-\theta_1(z-\lam+1/2) &  \theta_1(-z-\lam +1/2) \\  
\end{array} 
\right)
\ee
Some essential part of the vertex-IRF correspondence \cite{Baxter2} is 
expressed in terms of the matrices  
 \be
 S(w, \lam)^{(2)} \, S(z, \lambda - 2\eta h^{(2)})^{(1)} \, R(z-w, \lam) 
= R_{8V}(z-w)\,  S(z, \lam)^{(1)} S(w, \lam-2\eta h^{(1)})^{(2)} \, . 
\label{vertex-IRF} 
\ee
Here the symbol $S(z, \lam)^{(j)}$ denotes the matrix $S(z, \lam)$ 
acting on the $j$th space $V(z_j)$ of the tensor product $W=V^{\otimes L}$.  
 Next, we consider  the tensor product of the matrices acting on $W$ 
\be 
S_L(\lam)=S(z_L, \lam)^{(L)} S(z_{L-1}, \lam-2\eta h^{(L)})^{(L-1)} \cdots 
S(z_1, \lam - 2 \eta \sum_{j=2}^{L} h^{(j)})^{(1)}   
\label{matrixSn} 
\ee 
Then, through the vertex-IRF correspondence \cite{Baxter2,FV2}, we have  
\be L_{8V}(z) S(z, \lam)^{(0)} S_L(\lam-2\eta h^{(0)})   
= S_L(\lam) S(z, \lam-2\eta h)^{(0)} L(z, \lam) , 
\ee 
where $h= \sum_{j=1}^{L} h^{(j)}$. 
Let us denote by the symbol $u_c$ the eigenvector $b(t_1) \cdots b(t_m)v_c$  
constructed in \S 3 . 
Here we assume that the rapidities $t_1, \ldots, t_m$ satisfy the 
Bethe ansatz equations (\ref{BAE}). 
We also note that $h \, u_c = (L - 2m) \, u_c = r N \, u_c $. When 
$r m_1$ is even, we can show       
\be 
T_{8V}(z) S_L(\lam) \, u_c = S_L(\lam+ 2\eta) a(z, \lam+2\eta) \, u_c 
  + S_L(\lam-2\eta) d(z, \lam-2\eta) \, u_c \, . \label{shift}
\ee  
 Let us now define the following 
 \be 
 \Phi_c(\lam) = \sum_{k=0}^{2N-1}S_L(\lam+2\eta k) u_c(\lam+2\eta k) \, , 
\label{sum}
 \ee
 where  the parameter $c$ satisfies the condition: $\exp( 4 N \eta c )=1$. 
 Here we note that in the construction (\ref{sum}) 
 of $\Phi_c$, we have assumed that $u_c(\lam + 4N \eta c)= u_c(\lam)$.     
It follows from (\ref{shift})  that $\Phi_c(\lam)$ gives 
an eigenvector of the transfer matrix $T_{8V}(z)$ with the eigenvalue 
$C_0(z)$ defined by eq. (\ref{eigenvalue}):  
 $ T_{8V}(z) \Phi_c  = C_0(z) \, \Phi_c $. 
 Thus, we have discussed the construction of  eigenvectors for the discrete 
 $\eta$ case of $2 N \eta= m_1$,  
  when $L - 2 m = r N \, (r \in {\bf Z})$ and $r m_1$ is even. 
 Here we recall $W=V^{\otimes L}$ and $\Lam=n=L$.

\par 
Let us discuss the possible connection of the present method for 
constructing eigenvectors of the eight-vertex model to that of Ref. \cite{Baxter1,Baxter2,Baxter3}.  
In fact, it is not clear yet how 
 Baxter's method for  the the general case discussed in Refs. \cite{Baxter2,Baxter3} 
 is  related to the present method explicitly.  
   We first note that the parameter $\lam$ in the paper 
should  correspond to the parameter $s$ or $t$ of  Baxter's eigenvectors 
 in Refs. \cite{Baxter1,Baxter2,Baxter3}. 
 However, it seems that there is a large difference between the ways  
 how to  control the parameter $\lam$ or $s$  for the two methods. 
Thus, although there could exist more intrinsic 
connections between the two methods,     
we  can only point out here that  some important relations  
are common to both of them as mathematical formulas, 
such as the vertex-IRF correspondence (\ref{vertex-IRF}) 
and the sum  over the eigenvectors in eq. (\ref{sum}).     
Furthermore, we can compare some examples of eigenvectors constructed 
by the two methods, explicitly. For instance,  the eigenvector constructed in Ref. \cite{Baxter1} 
corresponds to the eigenvector $\Phi_c$ with $m=0$ and $c=0$ 
for the case of $2N\eta = 2m_1 + 2 m_2 \tau$ in the paper.

  \setcounter{equation}{0} 
  \renewcommand{\theequation}{5.\arabic{equation}} 
\section{Degenerate eigenvectors}

Let us construct degenerate eigenvectors 
for the transfer matrix of the eight-vertex model. 
 We can also construct eigenvectors  
 of  the 8VSOS, ABF and cyclic SOS models similarly by the method. 
 First, we recall that the number $m$ of the $b$ operators 
in the product $b(t_1) \cdots b(t_m)v_c $  
satisfies the condition (\ref{rN}): $L - 2m = r N$.   Here we note 
that $r m_1$ is even for the eight-vertex model,   
and also that $r=0$ for the 8VSOS and ABF models.

\par 
Let us now assume that out of $m$ rapidities $t_1, \ldots, t_m$, 
the first $R$ rapidities $t_j$ for $j=1, \ldots, R$ are of standard ones 
 satisfying the Bethe ansatz equations (\ref{BAE}) with $m$ replaced by $R$,  
while  the remaining $N F$ rapidities are
  formal solutions given by  
\be 
t_{(\alpha, j)} = t_{(\alpha)} + \eta(2j-N-1) + 
\epsilon_{\al} r_{j}^{(\alpha)}  \, , 
\qquad {\rm for} \quad j=1, \ldots, N\, . 
\ee 
Here $\epsilon_{\al}$'s are free parameters, and   
we shall consider  the limit of sending  $\epsilon_{\al}$  to zero, later. 
We call the set of $N$ rapidities $t_{(\alpha, 1)}, \ldots, t_{(\alpha, N)}$, 
the complete $N$-string with  center $t_{(\alpha)}$. Here   
 the index $\al$ runs from 1 to $F$. 
Furthermore, we  assume that the index $(\al, j)$ corresponds to the number 
$R+ N (\al-1) +j$ for $1 \le \al \le F$ and $1 \le j \le N$. 
  Here we note that  the importance of complete $N$-strings has been recently discussed  
for the $sl_2$ loop algebra symmetry of the six-vertex model  \cite{FM}, 
while  for the eight-vertex model   
the complete strings have been suggested 
in Ref. \cite{Baxter3} briefly in another context.

\par 
When $\epsilon_{\alpha}$'s are not zero, through the commutation relations such as shown in eq. (\ref{fcr}),
we can show the following relation  
\bea 
T(z)b(t_1) \cdots b(t_{R+NF}) \, v_c & = &
C_0(z) b(t_1) \cdots b(t_{R+NF})\, v_c \non \\
& + & \left( \sum_{j=1}^{R} +\sum_{j=R+1}^{R+NF} \right) \, C_{j} \, b(t_1) \cdots b(t_{j-1}) b(z) b(t_{j+1}) 
\cdots b(t_{R+NF}) \, v_c  \, . \label{Cbbb}
\eea
Now, let us consider the limit of sending $\epsilon_{\alpha}$ to zero. 
If we naively take the zero limits for $\epsilon_{\al}$'s, then  
the convergence of  the R.H.S. of eq. (\ref{Cbbb}) 
is not certain. We specify the limit as follows: 
setting $\epsilon_{\alpha}=\epsilon$, we  
divide  eq. (\ref{Cbbb}) by $\epsilon^F$, 
  and send $\epsilon$ to zero.  
Then, we can show that each of the terms 
 of eq. (\ref{Cbbb}) indeed converges, 
by making use of the following formula 
\be 
 \prod_{1 \le \al< \beta \le m} f_{P \al P \beta } 
= \prod_{1 \le \al< \beta \le m } f_{\al \beta} \, 
 \, \times \, 
\prod_{1 \le j < k \le m} 
 \left(   {\frac {\theta(t_j-t_k + 2\eta)}{\theta(t_j-t_k - 2\eta)}}
  \right)^{H(P^{-1}j - P^{-1}k)}  , {\rm for} \quad P \in {\cal S}_m 
\label{formula} 
\ee
Here  $H(x)$ denotes the Heaviside step function: $H(x)=1$ for $x> 0$,  
$H(x) = 0$ otherwise. The symbol  $P \in {\cal S}_m$ 
denotes an element $P$ of the  symmetric group of $m$ elements,  
where $j$ is sent to  $Pj \in \{ 1, 2, \ldots, m \}$ for $j=1, \ldots m$.  We recall that 
$f_{jk} = \theta(t_j - t_k - 2\eta) / \theta(t_j-t_k)$.   
The formula (\ref{formula}) has been  proven in Ref. \cite{XXX}. 

\par 
Let us consider  
an explicit formula describing the ``matrix elements'' of $b(t_1)\cdots b(t_m)v_c$, 
 which has been derived in Ref. \cite{FV2} for the case of $\Lam =2m$ and generic $\eta$. 
We can show that almost the same formula is valid also for the 
case of $\Lam -2m \ne 0$ when $2N \eta = m_1 + m_2 \tau$. 
For the tensor product $W=V^{\otimes L}$, we have 
\bea
& & b(t_1) \cdots b(t_m) v_c  =  (-1)^{m} e^{c(\lam + 2\eta m)} 
 \sum_{P \in {\cal S}_m} \sum_{1 \le j_1 < \cdots < j_m \le L} 
  \left( \prod_{\al=1}^{m} \prod_{\beta =j_{\al}+1}^{L} 
  {\frac {\theta(t_{P\al} - z_{\beta})} {\theta(t_{P\al} - z_{\beta} - 2 \eta)} }  
  \right) \non \\
&& \times  \, \prod_{1 \le \al < \beta \le m} f_{P\al P\beta} \, \times \,  
\prod_{\al=1}^{m} 
{\frac 
{\theta(\lam + t_{P \al} - z_{j_\al} - 2\eta(L-2m -j_{\al} + \al))}  
{\theta(t_{P\al} - z_{j_{\al}} - 2 \eta )} } 
\, \sigma^{-}_{j_1} \cdots \sigma^{-}_{j_m} \, |0 \ra 
\label{bbb}
\eea
Here $\sigma_{j}^{-}$ denotes the Pauli matrix $\sigma^{-}$ acting on 
the $j$th vector space, and $| 0 \ra$ the vacuum vector with all spins up. 
Here we also note that $p_k = z_k$ and $q_k= z_k + 2\eta$ 
since $\Lam_k =1$ for $k = 1, \ldots, L$.

\par 
We now discuss the behaviors of the terms of  
 eq. (\ref{Cbbb}) under the limit of $\epsilon$ sent to zero. 
In the limit, the factors $\prod_{1 \le \al < \beta \le m} f_{P\al P\beta}$ 
play the most important role in eq. (\ref{bbb}).  
With the formula (\ref{formula}), 
we can single out  the vanishing factors from the expression.   
We can show that the second part of (\ref{formula})  converges in the limit, 
which is given by  
$\prod_{1 \le j < k \le m} 
 \left(  {\theta(t_j-t_k  + 2\eta)}/{\theta(t_j-t_k - 2\eta)}
  \right)^{H(P^{-1}j - P^{-1}k)}$.  
For the complete strings $t_{(\al, j)}$  with $ 1 \le j \le N$, 
we have specified their ordering 
such that $t_{(\al, j)}-t_{(\al,k)}= 2\eta(j-k) + O(\epsilon)$ 
for $1 \le j < k \le N $.  
Thus, the factor ${\theta(t_{(\al, j)}-t_{(\al, k)} + 2\eta)}/
{\theta(t_{(\al, j)}-t_{(\al, k)} - 2\eta)}$ for an  
ordered pair of $j <k$  never diverges except when $j=1$ and $k=N$. 
If $P^{-1}(\al, 1) > P^{-1}(\al, N) $, then we have  ordered pairs 
of $(\al, j)$ and $(\al, j+1)$ for some $j$'s 
such that $P^{-1}(\al, j) > P^{-1}(\al, j+1)$, 
whose factors  cancel out the vanishing denominator: $\theta(t_1-t_N - 2\eta)$.  
Here we note that $t_{(\al, j)} - t_{(\al, j+1)} + 2 \eta = O(\epsilon)$. 

\par 
Let us denote by  $r_{a,b}^{(\al)}$ the difference: 
 $r_{a,b}^{(\al)}= r_{a}^{(\al)} - r_{b}^{(\al)}$.  
We can show that all the terms of R.H.S. of eq. (\ref{Cbbb}) 
converge under the limit $\epsilon \rightarrow 0$, 
if $r_{a,b}^{(\alpha)}$'s satisfy  the following   
\be 
{\frac {r_{a-1, a}^{(\al)}} {r_{a,a+1}^{(\al)}}}
= (-1)^{r m_1} e^{-4\eta c} 
\prod_{k=1}^{R}  
{\frac {\theta(t_{(\al,a)} - t_k + 2 \eta)} {\theta(t_{(\al,a)} -t_k - 2\eta )} } 
\prod_{\beta=1}^{L} 
{\frac {\theta(t_{(\alpha,a)}- z_{\beta}-2 \eta )} 
 {\theta(t_{(\alpha,a)}- z_{\beta})} } \, , \quad {\rm for} \quad 
 a=1, \ldots, N\, .  
\label{ratio}
\ee
Here we have assumed for $a=N$ and $a=1$ 
 the L.H.S. of (\ref{ratio})
denotes  $r_{N-1, N}^{(\al)} / r_{N,1}^{(\al)}$ 
and $r_{N, 1}^{(\al)}/r_{1,2}^{(\al)}$, respectively.  
 The center $t_{(\al)}$  of the complete $N$-string should  
satisfy the two constraints on $r_{a,b}^{(\alpha)}$'s 
 that are expressed in terms of the center $t_{(\al)}$
through eq. (\ref{ratio}), as follows    
\bea & & {\frac {r_{1,2}^{(\al)}} {r_{N,1}^{(\al)}} }
  + {\frac {r_{2,3}^{(\al)}} {r_{N,1}^{(\al)}} }+ \cdots 
+ {\frac {r_{N-1,N}^{(\al)}} {r_{N,1}^{(\al)}}} + 1 =0 \, , \label{rrr} \\
& & 
{\frac {r_{1,2}^{(\al)}} {r_{2,3}^{(\al)}} } \cdot 
   {\frac {r_{2,3}^{(\al)}} {r_{3,4}^{(\al)}} } \cdots 
 {\frac {r_{N-1,N}^{(\al)}} {r_{N,1}^{(\al)}}} \cdot 
 {\frac {r_{N,1}^{(\al)}} {r_{1,2}^{(\al)}}} = 1 \, . \label{rrr2} 
\eea   
It is easy to see from eq. (\ref{ratio}) 
that the constraint (\ref{rrr2}) holds if and only if 
$\exp(4N \eta \,c)= 1$.  For the eight-vertex model, the condition has  already been 
considered when we construct $\Phi_c$ in \S 4. 
Therefore,  eq. (\ref{rrr}) gives the only constraint  on the centers $t_{(\al)}$'s.  
 From solutions of eq. (\ref{rrr})   
 we can construct degenerate  eigenvectors 
 with the eigenvalue $(-1)^{F m_1} C_0(z)$ with the rapidities $t_1 \ldots, t_R$, 
for any $N$th root of unity assigned to  $\exp(4 \eta \,c)$.

\par 
We now discuss solutions of the eq. (\ref{rrr}). 
Let us introduce a function of variable $z$  
\bea 
G(z) & = &  
1 + \sum_{a=1}^{N-1}
 (-1)^{r m_1 a} e^{-4 \eta c a}  \non \\ 
 & & \times   \, \prod_{j=a+1}^{N} 
 \left(\prod_{k=1}^{R} {\frac {\theta(z-t_k + \eta(2j -N+1) )} {\theta(z-t_k + \eta(2j -N-3) )} } 
\prod_{\beta=1}^{L} {\frac {\theta(z- z_{\beta} + \eta(2j -N-3) )} 
 {\theta(z- z_{\beta} + \eta(2j -N-1))} } \right) 
\eea
Then, the equation  (\ref{rrr}) of centers $t_{(\al)}$ 's  is expressed as 
 \be 
 G(z=t_{(\al)})= 0 \,  , \quad {\rm for } \quad \al=1, \ldots, F \, . 
\label{GGG} 
 \ee 
The function $G(z)$  
is an elliptic function of $z$ with periods 1 and $\tau$ when $r m_1$ is even, and 
with periods 1 and $2\tau$ when $rm_1$ is odd.   
 Making use of the Bethe ansatz equations (\ref{BAE}) for $R$ rapidities  $t_1 \ldots, t_{R}$,    
 we can show that $G(z)$ has exactly $L$ poles. It follows from the theorem of  elliptic functions  
 that $G(z)$ has  $L$ zeros, which give solutions of centers $t_{(\al)}$'s. 
For an illustration,  the function $G(z)$ for the case of $N=3$  is given by   
\be 
G(z)  =  1 + f(z) + {\frac 1 {f(z+4\eta)}}
\label{Gf}
\ee
where $f(z)$, which corresponds to  $r_{1,2}^{(\al)}/r_{3,1}^{(\al)}$,  is explicitly  given by 
\be 
f(z) = (-1)^{rm_1} e^{4 \eta c} \prod_{k=1}^{R} 
{\frac {\theta(z- t_{k}- 4\eta)} {\theta(z- t_{k})}}
\prod_{\beta=1}^{L} {\frac {\theta(z - z_{\beta} -2\eta)}   
{\theta(z - z_{\beta} -4\eta)}}
\ee

\par 
Let us now discuss the dimensions of degenerate eigenspaces.  
In fact, the $L$ zeros of $G(z)$  do  not necessarily  give 
independent eigenvectors, as we shall see shortly. 
For an illustration, let us consider the case of 
$R=0$, explicitly.  Here we also assume that $L/N$ is an even integer. 
We can show that if $G(z)=0 $, then  we  have $G(z + 2\eta)=0$. 
For example, let us consider  the case of $N=3$. 
If $G(w_1)=0$, then we have $f(w_1+2\eta)f(w_1+4\eta)G(w_1)=0$.  
 From eq. (\ref{Gf}) and  $f(z)f(z+2\eta)f(z+ 4\eta)=1$, we have  
\bea 
f(w_1+2\eta)f(w_1+4\eta)G(w_1) & = & f(w_1)f(w_1+2\eta)f(w_1+4\eta) 
+ f(w_1+2\eta)f(w_1+4\eta)+f(w_1+2\eta) \non \\
&  = & 1 +  {\frac 1 {f(w_1)}} + f(w_1+2\eta)  \non \\
& = & G(w_1+ 2\eta ) \, .  
\eea
We thus have a set of $N$ zeros of $G(z)$ as $w_1, w_1 + 2\eta, \ldots, w_1 + 2 (N-1) \eta$.  
The $L$ zeros of $G(z)$ is given by $L/N$ sets of complete $N$-strings with $L/N$ centers.  
Therefore, the number of independent solutions is given by $L/N$. 
Here we note that the $L/N$ centers  should be independent in general, 
since the inhomogeneous parameters $z_{\beta}$'s are generic.  

\par 
Let us show that the number $2^{L/N}$ gives  
the dimension of the degenerate eigenspace of the XYZ model 
thus constructed for $R=0$ with an  $N$th root of unity assigned to  $\exp(4 \eta c)$.   
 We can calculate it by  the binomial expansion 
\be 
2^{L/N} = \sum_{F=0}^{L/N}  {\frac {(L/N)!} {(L/N-F)! F!}} \, , 
\ee
which corresponds to  the sum of the dimensions 
of the degenerate subspaces  with $F=0, 1, \ldots$, and $L/N$, respectively.     
Here,  the dimension of the degenerate sub-eigenspace with $F$ sets of complete $N$-strings 
is given by the number of ways for selecting $F$ centers 
from the $L/N$ centers of the $L/N$ complete $N$-strings.  
Here we note that the factor $(-1)^{Fm_1}$ for the eigenvalue of the transfer matrix 
is canceled when we take the logarithmic derivative of the eigenvalue of 
the transfer matrix as shown in the eq. (125) of Ref. \cite{Baxter2}.

\par 
Let us now remark that  we have an extra factor $N$ for the dimension of the 
degenerate eigenspace of the XYZ spin chain with $R=0$, 
 since   we may choose any $N$th root of unity for $\exp(4 \eta c)$.  
Thus, we  obtain the number $N \, 2^{L/N}$ as the degeneracy in total. 
In fact, we can show that the degenerate eigenvalue 
of the XYZ Hamiltonian with $R=0$  does not 
depend on the choice of $c$ by using  eq. (125) of Ref. \cite{Baxter2}, 
or more explicitly, by  eq. (5.37) of Ref. \cite{8VABA}.  
 For the case of $L=12$ and $N=3$, the numbers  $2^{L/N}$ and $N 2^{L/N}$ 
 are given by 16 and 48, respectively,  which are consistent 
 with  the numerical result \cite{Klaus}.

\par Let us now consider the trigonometric limit for the degenerate eigenvectors 
of the XYZ spin chain. After taking the limit, the number of 
down spins $M$ becomes a good quantum number for the eigenvectors.  
The limit of the eigenvector $\Phi_c$ of the XYZ model corresponds to  a linear combination 
 of eigenvectors of the XXZ model labeled by several $M$'s. 
 Here we note that we replace $\lam$ with $\lam+\tau/4$, and also that  
 $\theta_1(z+\tau/2) \rightarrow i \exp(-\pi z)$ and 
 $\theta_0(z+ \tau/2) \rightarrow 1$  when $\tau \rightarrow i \infty$. 
 Thus, from the expression (\ref{matrixSn}) 
of $S_L(z, \lam)$, we can show that 
the matrix elements of $S_L(z, \lam)$ in the sector of  $M$ down-spins 
have the factor $\exp(- \pi i M \lam)$ with respect to the variable $\lam$.  
Since the eigenvector $\Phi_c$ is given by the product of the matrix $S_L(z, \lam)$ and 
the vector $b(t_1) \cdots b(t_m) v_c$,  all the entries of the XXZ eigenvectors 
 with $M$ down-spins derived from $\Phi_c$ have the factor 
 $\exp( (c - \pi i (m+M)) \lam ) $ with respect to the variable $\lam$ .  
We now recall  the sum  (\ref{sum}). 
If $\exp(2\eta (c - \pi i (m+M) ) ) \ne 1$, 
then the XXZ eigenvectors  derived from the XYZ eigenvector $\Phi_c$ 
do not have $M$ down-spins.   Here we note 
\be 
\sum_{j=0}^{2N-1} \exp \left( 2 \eta j (c - \pi i (m + M) )  \right) = 0 \, . 
\ee
When $\exp(2\eta (c - \pi i (m+M) )) = 1$, the XXZ eigenvectors with $M$ down-spins 
derived from  $\Phi_c$  do not vanish.  
 For the degenerate eigenspace with the dimension $2^{L/N}$ for $R=0$ and $c=0$,  
 we put $M=m$ in  each sector of $m$, 
where $m$ is given by $m=NF$ for $F = 0, \ldots, L/N$. 
Thus, the dimension of the degenerate XXZ eigenvectors derived from 
the XYZ eigenvector with $c=0$ through the trigonometric limit 
is given by $2^{L/N}$, in total. We note that it is equivalent to 
the dimension of the largest degenerate eigenspace  associated with  
the $sl_2$ loop algebra  of the XXZ spin chain with $L$ lattice sites 
in the sector $S^{Z} \equiv 0 \, ({\rm mod} \, N)$ \cite{DFM}, 
where  $S^{Z}=(L-2M)/2$. Furthermore, we can explicitly calculate by using  eq. (\ref{bbb})   
 the trigonometric limits of the degenerate eigenvectors of the XYZ spin chain 
with the complete $N$-strings. Let us  consider the case of  $R=0$ and $c=0$:    
%
%
 when $F=0$ or $L/N$,  the limits 
correspond to the degenerate XXZ eigenvectors associated with the $sl_2$ loop algebra \cite{DFM};  
 when $F =1, \ldots, L/N-1$, however, 
we have a conjecture that the limits 
 can  be expressed as linear combinations of the degenerate XXZ eigenvectors  
 of the complete $N$-strings,  which have been shown to be associated with 
 the $sl_2$ loop algebra in Refs. \cite{FM}. Thus, as a summary of the above discussion, 
 we may conclude that at least some spectral degeneracy of the XYZ model is related to 
that of the XXZ model associated with the $sl_2$ loop algebra.

\section{Discussions} 

\par 
We have shown that there exists a large degeneracy in the spectrum of the XYZ spin chain 
for the discrete $\eta$ case of $2 N \eta = m_1 + m_2 \tau$. 
The dimension $N 2^{L/N}$ of the degenerate eigenspace  
increases exponentially with respect to the lattice length $L$. Thus, 
we may have a conjecture that the spectral degeneracy of the XYZ model 
 should have nontrivial effects on the continuum limits.  
We also note that in  the connection between  the spectra 
of the eight-vertex model and  the six-vertex  model under the  trigonometric limit,   
the Bethe ansatz equations for the XXZ model under the twisted  
boundary conditions should play a central role. We  shall discuss the connection 
explicitly in later publications.

\par 
The relation (\ref{R-w}) between the $R$-matrix of $E_{\tau,\eta}(sl_2)$ and 
that of the 8VSOS model is fundamental in the elliptic quantum groups. 
We may interpret that the $R$-matrix of $E_{\tau,\eta}(sl_2)$ is  derived 
from that of the 8VSOS model through the relation. 
In fact, we can discuss the $R$ matrices of the elliptic quantum groups 
associated with the ABCD-type Lie algebras \cite{Felder} 
based on  the Boltzmann weights \cite{A-IRF,ABCD} of the ABCD IRF models. 

\par  
As an application of the observation in the last paragraph, 
we can discuss the spectral degeneracy 
of the $A$-type IRF models \cite{A-IRF} through 
the nested algebraic Bethe ansatz of the associated 
elliptic quantum groups. Here we note that the  $A$-type  IRF models 
are related to the elliptic vertex  model  \cite{Belavin},   
which corresponds to the higher-rank extension of the eight-vertex model,  
through the vertex-IRF correspondence.  
This approach should be related to the higher-rank loop-algebra symmetry 
of the vertex models discussed in Ref. \cite{Korff}. 

\par Finally, we remark that it should be  straightforward to discuss 
the spectral degeneracy for the higher-spin generalization of the XYZ model  
by making use of the higher-spin version  \cite{Takebe} of the vertex-IRF correspondence.

{\vskip 1.2cm}
\par \noindent 
{\bf Acknowledgements }

The author would like to thank  Prof. K. Fabricius and Prof. B.M. McCoy 
for helpful discussions and  valuable comments.  
 He would also like to thank  Prof. M. Kashiwara and  Prof. T. Miwa 
 for their kind invitation to  the conference ``MathPhys Odyssey 2001'', Okayama-Kyoto, February, 2001, 
where quite an early version of this paper has been presented. He  
 is  thankful to many participants of the conference for their helpful comments. 
This work is partially supported by the Grant-in-Aid for Encouragement 
of Young Scientists (No. 12740231).

\newpage 


\setcounter{section}{0}
\renewcommand{\thesection}{\Alph{section}}



\setcounter{equation}{0} 
 \renewcommand{\theequation}{A.\arabic{equation}}
\section{Appendix: The $sl_2$ loop algebra }

Let us introduce the loop algebra for $sl(2, {\bf C})$. 
The loop algebra associated to $sl_2$ consists 
of the space of analytic mappings from the circle $S^1$ 
to $sl_2$. If $T^a$ is a basis of $sl_2$, 
and $S^1$ is considered as the unit circle 
in the complex plane with coordinate $z=\exp(2\pi i t)$, 
then Fourier analysis shows that a topological basis 
of the vector space of these maps  is given by 
$T^a \otimes z^n$ for $n \in {\bf Z}$ \cite{Fuchs}. 
The commutation relation is defined by 
$[ T^a \otimes z^m , T^b \otimes z^n ] = [ T^a, T^b] \otimes z^{m+n}$.   
The affine Lie algebra  $\hat{sl}_2$  
is given by the central extension 
of the $sl_2$ loop algebra together with the derivation $d$ :  
$\hat{sl}_2  =  sl_2 \otimes {\bf C}[z, z^{-1}] \oplus {\bf C} c 
\oplus {\bf C} d$, where the commutation relations are given by  
\bea 
{[} T^a \otimes z^m , T^b \otimes z^n {]} &  = & 
{\rm [} T^a, T^b {\rm ]} \otimes z^{m+n} +
 m \delta_{m+n, 0} \, {\rm tr} (T^a T^b) c \, , \non \\
{\rm [} d,  T^a \otimes z^m {\rm ]} & = & m T^a \otimes z^m  \, , \qquad  
{\rm [} c,  T^a \otimes z^m {\rm ]}  = 0 \, .   
\eea  
We shall also consider the affine algebra 
 with no derivation $d$: $\hat{sl}_2^{'} =  
sl_2 \otimes {\bf C}[z, z^{-1}] \oplus {\bf C} c$.   

\par 
Let us  denote the Chevalley generators of the $sl_2$
by  $e$, $f$ and $h$.  Then, 
we may define the Chevalley generators of $\hat{sl}_2$ by the following:  
 $h_0 = c- h \otimes 1$, $h_1 = h \otimes 1$, 
 $e_0= f \otimes z$,  $f_0= e \otimes z^{-1}$,  
 $e_1= e \otimes 1$,  $f_1= f \otimes 1$ and $d=t \partial/\partial t$.     
The Cartan subalgebra ${\hat {\bf h}}$ of $\hat{sl}_2$ is spanned 
by $h_0$, $h_1$ and  $d$.  
In terms of the Chevalley basis, 
the defining relations of $\hat{sl}_2$ are given by 
\bea 
{\rm [} h_0, h_1 {\rm ]} & = & 0 \, , \quad 
 {\rm [} d, h_j {\rm ]} =0   \qquad ( j =0, 1 ) \non \\
{\rm [} h_i, e_j {\rm ]}  & = & a_{ij} e_j \, ,  \quad 
 {\rm [} h_i, f_j {\rm ]}   = - a_{ij} f_j \, ,   \quad 
 {\rm [} e_i, f_j {\rm ]}  =  \delta_{ij} h_{j} \, , 
  \qquad (i,j = 0, 1) \non \\
{\rm [} d, e_j {\rm ]}  & = & \delta_{0,j} \,  e_j \, ,  \quad 
 {\rm [} d, f_j {\rm ]}   = - \delta_{0j} \, f_j \, ,   \qquad 
 ( j =0, 1 )
\non \\
{\rm [} e_i,  {\rm [} e_i, {\rm [} e_i,  e_j {\rm ]}   {\rm ]}   {\rm ]} & =& 0    
\, , \quad 
{\rm [} f_i,  {\rm [} f_i, {\rm [} f_i,  f_j {\rm ]}   {\rm ]} =0   
\qquad (i,j=0,1, \quad  i \ne j) 
\label{Chevalley} 
\eea 
Here, the Cartan matrix $(a_{ij})$ of $A_1^{(1)}$ is defined  by 
\be 
\left( 
\begin{array}{cc}  
a_{00} & a_{01} \\
a_{10} & a_{11} 
\end{array} 
\right) 
 = \left( 
\begin{array}{cc} 
2 & -2 \\
-2 & 2 
\end{array}
\right)
\ee

\par   
Let us now review the connection of the $sl_2$ loop algebra 
to the XXZ Hamiltonian $H_{XXZ}$ at the root of unity.  
Hereafter, we assume that 
$q^{2N}=1$.   
We introduce the operators $S^{\pm(N)}$ by  
\begin{eqnarray}
S^{\pm(N)}&=&  
\sum_{1 \le j_1 < \cdots < j_N \le L}
q^{{N \over 2 } \sigma^Z} \otimes \cdots \otimes q^{{N \over 2} \sigma^Z}
\otimes \sigma_{j_1}^{\pm} \otimes
q^{{(N-2) \over 2} \sigma^Z} \otimes  \cdots \otimes q^{{(N-2) \over 2}
\sigma^Z}
\nonumber \\
 & & \otimes \sigma_{j_2}^{\pm} \otimes q^{{(N-4) \over 2} \sigma^Z} \otimes
\cdots
\otimes \sigma^{\pm}_{j_N} \otimes q^{-{N \over 2} \sigma^Z} \otimes \cdots
\otimes q^{-{N \over 2} \sigma^Z} \quad . 
\label{sn}
\end{eqnarray}
We denote by  $T^{\pm(N)}$  the operators  obtained from $S^{\pm(N)}$ 
via  replacing  $q$ with  $q^{-1}$.   
Explicitly, we have 
\begin{eqnarray}
T^{\pm(N)} & =&  
\sum_{1 \le j_1 < \cdots < j_N \le L}
q^{- {N \over 2 } \sigma^Z} \otimes \cdots \otimes q^{- {N \over 2} \sigma^Z}
\otimes \sigma_{j_1}^{\pm} \otimes
q^{- {(N-2) \over 2} \sigma^Z} \otimes  \cdots \otimes q^{ - {(N-2) \over 2}
\sigma^Z}
\nonumber \\
 & & \otimes \sigma_{j_2}^{\pm} \otimes q^{- {(N-4) \over 2} \sigma^Z} \otimes
\cdots
\otimes \sigma^{\pm}_{j_N} \otimes q^{{N \over 2} \sigma^Z} \otimes \cdots
\otimes q^{{N \over 2} \sigma^Z} \quad . 
\label{tn}
\end{eqnarray}
Let the symbol $T_{6V}(v)$ denotes the (inhomogeneous) transfer matrix 
of the six-vertex model.    
Here we recall that $S^Z$ denotes the $Z$-component of the total spin operator. 
Then we can show the (anti) commutation relations 
in the sector of $S^Z \equiv 0~ (\rm{mod}~N)$ \cite{DFM} 
\begin{equation}
S^{\pm (N)} T_{6V}(v)=q^N T_{6V}(v) S^{\pm (N)}, \qquad 
T^{\pm (N)} T_{6V}(v)=q^N T_{6V}(v) T^{\pm (N)} 
\end{equation}
Since the XXZ Hamiltonian $H_{XXZ}$ is given by 
the logarithmic derivative of the (homogeneous) transfer matrix 
$T_{6V}(v)$,    
we have  in the sector $S^Z \equiv 0$ (mod $N$)  
\begin{eqnarray}
{[}S^{\pm(N)},H_{XXZ} {]}={[}T^{\pm(N)},H_{XXZ} {]}=0.
\label{sthcomm}
\end{eqnarray}
Thus, the operators $S^{\pm (N)}$  and $T^{\pm (N)}$ commute 
with the XXZ Hamiltonian in the sector  $S^Z \equiv 0$ (mod $N$) .  

\par 
Let us  now consider the algebra generated by 
 the operators $S^{\pm (N)}$ and $T^{\pm (N)}$  \cite{DFM}. 
When $q$ is a primitive $2N$th root of unity, 
or a primitive $N$th root of unity with $N$ odd, 
we can show that $S^{\pm (N)}$ and $T^{\pm (N)}$  
satisfy  the relations in the following \cite{DFM}: 
\be
[S^{+(N)}, T^{+(N)}]  =  [S^{-(N)},T^{-(N)}] = 0 \, ,  \label{one}  
\ee
\be 
[S^{Z}, S^{\pm (N)}]  =  \pm N S^{ \pm (N)}, \qquad 
[S^{Z}, T^{\pm (N)}]=   \pm N T^{\pm (N)} \, , \label{anothertwo} 
\ee
\begin{eqnarray}
(S^{+(N)} )^3 \, T^{-(N)} - 3 (S^{+(N)})^2 \, 
T^{-(N)} \, S^{+(N)} + 3 S^{+(N)} \, T^{-(N)} \, (S^{+(N)})^2 
-T^{-(N)} \, (S^{+(N)})^3 &=& 0 \, , \non \\
(S^{-(N)})^3 \, T^{+(N)} - 3 (S^{-(N)})^2 \,T^{+(N)} \, S^{-(N)}
+ 3 S^{-(N)} \, T^{+(N)} \, (S^{-(N)})^2 
-T^{+(N)} \, (S^{-(N)})^3 &=&0 \, , \non \\
(T^{+(N)})^3 \, S^{-(N)} - 3 (T^{+(N)})^2 \, S^{-(N)} \, T^{+(N)}
+3 T^{+(N)}\, S^{-(N)} \, (T^{+(N)})^2 
- S^{-(N)} \, (T^{+(N)})^3 &=&0 \, , \non \\
(T^{-(N)})^3 \, S^{+(N)} -3 (T^{-(N)})^2 \, S^{+(N)} \, T^{-(N)}
+3 T^{-(N)} \, S^{+(N)} \, (T^{-(N)})^2 
-S^{+(N)} \, (T^{-(N)})^3 &=&0 \, , \non \\
\label{serre} 
\end{eqnarray}
and in  the sector $S^z\equiv 0 ({\rm mod}~N)$ we have 
\begin{equation}
[S^{+(N)},S^{-(N)}]=[T^{+(N)},T^{-(N)}]=-(-q)^N {2 \over N} S^z. \label{two}
\end{equation}
When $q$ is a primitive $2N$th root of unity with $N$ even, or 
 when $q$ is a primitive $N$th root of unity with $N$ odd, then, 
we consider the  identification in the following:   
\begin{equation}
e_0=S^{+(N)}, \quad f_0=S^{-(N)}, \quad e_1=T^{-(N)}, \quad 
f_1=T^{+(N)}, \quad h_0=-h_1= {\frac 2 N}  S^z \, .    
\end{equation}
We see that the operators $e_j, f_j, h_j$ for $j=0,1$,  
satisfy the defining relations  (\ref{Chevalley}) of the algebra 
$\hat{sl}_2^{'}$ with $c=0$.   
The relations (\ref{one}) and (\ref{two}) correspond to some relations 
of  (\ref{Chevalley}) in the following: 
${\rm [} h_i, e_j {\rm ]} = a_{ij} e_j$ ,   
$ {\rm [} h_i, f_j {\rm ]}   = - a_{ij} f_j$ and 
$ {\rm [} e_i, f_j {\rm ]}  =  \delta_{ij} h_{j}$.  
The relations (\ref{serre}) correspond to the Serre relations of 
 (\ref{Chevalley}). 
 Thus, they give a representation of the $sl_2$ loop algebra  
or a finite-dimensional representation of $\hat{sl}_2^{'}$.

\par 
When $q$ is a primitive $2N$th root of unity with $N$ odd,   
we may put as follows 
\begin{equation}
e_0= i S^{+(N)}, \quad f_0= i S^{-(N)}, \quad e_1= i T^{-(N)}, \quad 
f_1= i T^{+(N)}, \quad h_0=-h_1= {\frac 2 N}  S^z \,  .   
\end{equation}
We see that they give a representation of the $sl_2$ loop algebra  
or a finite-dimensional representation of $\hat{sl}_2^{'}$.

\par 
Finally we note that the loop algebras with higher ranks are discussed for 
some vertex models \cite{Korff}.

\setcounter{equation}{0} 
 \renewcommand{\theequation}{B.\arabic{equation}}
\section{Appendix: Evaluation modules of the elliptic quantum group }

\par 
We first define  a diagonalizable $\bf h$-module \cite{FV1,FV2}.  
Let $V$ be a module over the one-dimensional Lie algebra ${\bf h}$ 
with the generator $h$ such that $V$ is the direct sum 
of finite dimensional eigenspaces $V[\mu]$ of $h$, 
labeled by the eigenvalue $\mu$. 
We call such a module a diagonalizable ${\bf h}$-module. 

\par 
Let us now define a module over 
the elliptic quantum group $E_{\tau, \eta}(sl_2)$ \cite{FV1,FV2}. 
An $E_{\tau, \eta}(sl_2)$ module is a diagonalizable 
$\bf h$-module $V$ together with a meromorphic function $L(z,\lam)$ 
on ${\bf C} \otimes {\bf h}$ with values in 
${\rm End}({\bf C}^2 \otimes V)$  such that the dynamical Yang-Baxter 
relation (\ref{DYBR}) holds in 
${\rm End}({\bf C}^2 \otimes {\bf C}^2 \otimes V)$. 

\par 
When $V={\bf C}^2$, $L(z,\lam)=R(z-z_0, \lam)$ gives a module 
over $E_{\tau, \eta}(sl_2)$, which we call 
the fundamental representation with evaluation point $z_0$. 
More generally, for any pair of complex numbers, $\Lam, z$,  
we can define an evaluation Verma module $W_{\Lam}(z)$  as follows \cite{FV1}.  
Let $W_{\Lambda}(z)$ be an infinite dimensional complex vector space 
with a basis $e_k$ $(k \in {\bf Z}_{\ge 0})$. We define 
an action of $f(h)$ by 
\be 
f(h) e_k = f(\Lam - 2k) e_k \quad (k = 0, 1, \ldots)  
\ee
and that of the other generators by 
\bea
a(w, \lam) e_k  & =& {\frac {\theta(z-w+(\Lam +1 -2k) \eta) }
{\theta(z-w+ (\Lam +1) \eta) } } \, 
 {\frac {\theta(\lam + 2k \eta)} {\theta(\lam)}  }\, e_k \, , 
\non \\
b(w, \lam) e_k & =& - {\frac {\theta(- \lam + z-w+(\Lam -1 -2k) \eta) }
{\theta(z-w+ (\Lam +1) \eta) } } \, 
 {\frac {\theta(2 \eta)} {\theta(\lam)}  }\, e_{k+1}  \, , 
\non \\
c(w, \lam) e_k & =& - {\frac {\theta(- \lam - z + w+ (\Lam+1 -2k) \eta) }
{\theta(z-w+ (\Lam +1) \eta) } } \, 
 {\frac {\theta(2(\Lam +1 - k) \eta)} {\theta(\lam)}  } 
 \,  {\frac {\theta(2k \eta)} {\theta(2 \eta)}  } \, e_{k-1}  \, , \non \\
d(w, \lam) e_k  & =& {\frac {\theta(z-w+(- \Lam +1 + 2k) \eta) }
{\theta(z-w+ (\Lam +1) \eta) } } \, 
 {\frac {\theta(\lam - 2(\Lam - k)\eta)} {\theta(\lam)}  }\, e_k \, . 
\eea
It is shown in Ref. \cite{FV1} that
these formulas define an $E_{\tau, \eta}(sl_2)$-module structure 
on $W_{\Lam}(z)$; if $\Lam = n + (m + \ell \tau )/2\eta$, where 
  $n$, $m$ and $\ell$ are integers with $n \ge 0$, then the 
  space spanned by 
  $e_k$, $k > n$, is a submodule. 
  The quotient space $L_{\Lam}(z)$ is a module of dimension $n+1$.

\par 
The evaluation Verma module $W_{\Lam}(z)$ is a highest weight module 
with highest weight vector 
$e_0$ and highest weight $(\Lam, A(w,\lam), D(w,\lam))$, where 
 $A(w,\lam)=1$, and 
\be
 D(w,\lam) = {\frac {\theta(z-w+(-\Lam+1) \eta)} 
{\theta(z-w+(\Lam+1) \eta)}} \,  
 {\frac  {\theta(\lam-2 \Lam \eta)} {\theta(\lam)}}
\ee

\par 
Let us denote  the finite dimensional module $L_{\Lam}(z)$ by $V_{\Lam}(z)$ 
if $\Lam= n + (m+ \ell \tau) / 2\eta$ where $n$, $m$, $\ell$ are integers with 
$n \ge 0$, and the infinite dimensional module $W_{\lam}(z)$ if $\Lam$ does  not have this form. 
Then, the following proposition plays an important role in \S 3.   
\begin{prop}\cite{FV1} 
Let $W =V_{\Lam_1}(z_1) \otimes \cdots V_{\Lam_n}(z_n)$ 
be the tensor product of evaluation modules $W_{\Lam_j}(z_j)$'s, and let   
$\Lam=\Lam_1 + \Lam_2 + \cdots + \Lam_n$. 
 Then,  $W[\Lam] = {\bf C} v_0$, and   
 for every $z$ we have 
\be 
a(z)v_0 = A(z, \lam) v_0 \, , \quad c(z)v_0 = 0\, , \quad  d(z)v_0 = D(z,\lam) v_0  
\ee
where 
\be 
A(z,\lam)=1, \quad D(z, \lam) = 
{\frac {\theta(\lam-2\Lam)} {\theta(\lam)}} \prod_{j=1}^{n} 
{\frac {\theta(z-p_j)} {\theta(z-q_j)} } 
\ee
Here $p_j= z_j + \eta(-\Lam_j +1)$ and 
$q_j= z_j + \eta(\Lam_j +1)$ .
\end{prop}

\setcounter{equation}{0} 
 \renewcommand{\theequation}{C.\arabic{equation}}
\section{Appendix: The IRF models related to the eight-vertex model} 

 Let us consider a two-dimensional square lattice 
 with a spin variable $a_i$ associated to each site $i$ \cite{Kuniba}. 
 We shall call the $a_i$ a state and assume that 
 $a_i \in S$ with $S$ being a set of the states. The set $S$ is finite 
 for restricted Solid-on-Solid models (RSOS models), while 
 it is infinite for unrestricted Solid-on-Solid models (unrestricted SOS models).  
Here we note that unrestricted and restricted SOS models 
are also called Interaction-Round-a-Face models or 
IRF models, briefly 
\cite{Baxter-book}. 

\par 
Let us consider the case of RSOS models.  Let $s$ denote 
the number of elements in $S$. Consider a $s \times s$ matrix $C$ 
satisfying the following conditions \cite{Kuniba,Akutsu}: 

\par 
(i) $C_{ab} = C_{ba} = 0$ \, {\rm or} \,  1 
\par 
(ii) $C_{aa} =0$ 
\par 
(iii) For each $a \in S$, there should exist $b \in S$ such that $C_{ab} =1$ 

\par \noindent 
 For such choice of $C$, we impose a restriction 
that two states $a$ and $b$ can occupy the neighboring lattice sites 
if and only if $C_{ab}=1$. We call such a pair of the states $(a,b)$ admissible. 
 For the case of unrestricted models, 
 the infinite matrix $C$ satisfies the conditions (i), (ii), and (iii) with 
an infinite set $S$.

\par 
 For an illustration, let us consider 
 the restricted eight-vertex Solid-on-Solid model 
(the restricted 8V SOS model), which we also call the ABF model \cite{ABF}.  
For the $N$-state case, we have $S=\{1, 2, \ldots, N\}$. 
The nonzero matrix elements of $C$ are 
given by $C_{j,j+1}=C_{j+1,j}=1$ for $j=1,2, \ldots, N-1$;   
other matrix elements such as $C_{1, N}$ and $C_{N,1}$  are 
given by zero.

\par 
Let $a_i, a_j, a_k$, and $a_{\ell} $ be the four states assigned on the lattice 
sites $i, j, k$, and $\ell$ surrounding a face. 
Here $i,j,k,\ell$ are ordered counterclockwise from  the 
southwest corner.   We assume that an elementary configuration  is given by 
the configuration of the four spin variables  
around the face and the probability 
of having $a_i, a_j, a_k, a_{\ell}$ around the face 
 is  denoted  by the Boltzmann weight 
$w(a_i, a_j, a_k, a_{\ell}; z)$.  
Here the variable $z$ is called the spectral parameter. 
The model is called solvable 
if the Boltzmann weights 
satisfy the Yang-Baxter relations in the following:  
\bea
& & \sum_g w(a,b,g,f; z-w) w(f,g,d,e; z) w(g,b,c,d; w) \non \\ 
& = & \sum_g w(f,a,g,e; w) w(a,b,c,g; z) w(g,c,d,e; z-w) \, , 
\label{ybr}
\eea
where the summation of the variable $g$ is taken over all the admissible states.

\par 
Let us introduce  explicitly some solutions of the Yang-Baxter relations. 
We consider a slightly  generalized version of eq. (\ref{R-w}), 
where $\lam$ is given by $\lam= - 2\eta d - w_0$ with some parameter $w_0$.  
Then, we have the Boltzmann weights $w(a,b,c,d; z, w_0)$ 
of the unrestricted eight-vertex Solid-on-Solid  model 
(unrestricted 8V SOS model)  
\bea 
w(d+1,d+2,d+1,d; z, w_0 ) & = &  w(d,d-1,d-2,d-1; z ) = 1  \non \\ 
w(d-1,d,d+1,d; z, w_0 ) & = &   \alpha(z, - 2\eta d- w_0)   \non \\ 
w(d+1,d,d-1,d; z, w_0 ) & = &   \alpha(z,  2\eta d + w_0)   \non \\ 
w(d+1,d,d+1,d; z, w_0 ) & = &   \beta(z,  2\eta d + w_0)   \non \\ 
 w(d-1,d,d-1,d; z,w_0 ) & = &   \beta(z, - 2\eta d- w_0) 
\label{bw}
\eea 
The Boltzmann weights (\ref{bw}) 
satisfy the Yang-Baxter relations (\ref{ybr}).  

\par 
Let us introduce a gauge transformation   
\be 
w(a,b,c,d ; z) \rightarrow w(a,b,c,d; z) \, {\frac {g_c} {g_a} } \, . 
\label{gauge-t}
\ee
Then, we see that the transformed Boltzmann weights also satisfy 
the Yang-Baxter relations (\ref{ybr}). 
Setting $g_a = \exp(\pi i a/2) \, \sqrt{\theta(2 \eta a + w_0) } $ 
($a \in {\bf Z}$), applying (\ref{gauge-t}) to (\ref{bw}), 
and  multiplying the Boltzmann weights  by 
$\rho(z)= \theta(2\eta -z)/ \theta (2\eta)$,  
we  have the standard expressions of the Boltzmann weights 
such as in Refs. \cite{ABF,Pearce,Kuniba,Akutsu} in the following:   
\bea 
w(d+1,d+2,d+1,d; z, w_0 ) & = &  w(d,d-1,d-2,d-1; z ) = 
{\frac {\theta(2\eta -z)} {\theta(2\eta)}} 
  \non \\ 
w(d-1,d,d+1,d; z, w_0 ) & = &   w(d+1,d,d-1,d; z, w_0 ) \non \\
& = & {\frac {\theta(z)} {\theta(2\eta)}} \, 
{\frac {\sqrt{\theta(2\eta (d+1) + w_0 ) \theta(2\eta (d-1) + w_0)}} 
{\theta(2\eta d + w_0)} } \non \\ 
w(d+1,d,d+1,d; z, w_0 ) & = &  
{\frac {\theta(z+ 2\eta d + w_0)}  {\theta(2\eta d + w_0)}}    \non \\ 
 w(d-1,d,d-1,d; z,w_0 ) & = &   
 {\frac {\theta(z- 2\eta d - w_0)}  {\theta(2\eta d + w_0)}}    
\label{bw2}
\eea

\par 
Let us now consider the Boltzmann weights of the ABF and CSOS models. 
Here we assume that $2 N \eta = m_1$, where integer $m_1$ has no common 
divisor with  $N$. If we set $w_0=0$, 
then we have the Boltzmann weights of the ABF model. 
We can show that the Boltzmann weights (\ref{bw2}) 
satisfy the Yang-Baxter relations (\ref{ybr}) 
with the finite set: $S= \{1,2, \ldots, N \}$. 
If we set $w_0 \neq 0$, 
then we have the Boltzmann weights of the cyclic SOS model (CSOS model). 
We can show that the Boltzmann weights (\ref{bw2}) 
satisfy the Yang-Baxter relations 
with the finite set: $S= \{1,2, \ldots, N \}$ and the 
cyclic admissible conditions: $C_{1,N} =C_{N,1} = 1$. 
Here we note this  RSOS model is called the 
cyclic SOS model in Ref. \cite{Pearce}, 
the $A_{N-1}^{(1)}$ model in Ref. \cite{Kuniba}, 
and the periodic 8V SOS model in Ref. \cite{Akutsu} 
(See also the Appendix of Ref. \cite{Akutsu}). 



\newpage  
\begin{thebibliography}{[99]}
\bibitem{Baxter0} R. Baxter, Ann. Phys. {\bf 70} (1972) 193. 
\bibitem{Baxter1} R. Baxter, Ann. Phys. {\bf 76} (1973) 1. 
\bibitem{Baxter2} R. Baxter, Ann. Phys. {\bf 76} (1973) 25. 
\bibitem{Baxter3} R. Baxter, Ann. Phys. {\bf 76} (1973) 48. 
\bibitem{Baxter-book} R.J. Baxter, 
{\it Exactly Solved Models in Statistical Mechanics} (Academic Press, London, 1982).  
\bibitem{Jimbo-book} 
 {\it Yang-Baxter equation in integrable 
systems} ed. by M. Jimbo (World scientific, Singapore, 1990).  


\bibitem{8VABA} L. Takhtajan and L. Faddeev, 
 Russ. Math. Survey {\bf 34}(5) (1979) 11. 
\bibitem{Krinsky} J.D. Johnson, S. Krinsky and B.M. McCoy, Phys. Rev. A {\bf 8} (1973) 2526. 
\bibitem{Luther} A. Luther, Phys. Rev. B (1976) {\bf 14} (1976) 2153
\bibitem{Pokrovsky} S. Pokrovsky and A.M. Tsvelik, Sov. JETP {\bf 93} (1987) 2232.  
\bibitem{Affleck} I. Affleck, {\it Fields, Strings and Critical Phenomena}
ed. E. Brezin and J. Zinn-Justin (1990, Amsterdam, North-Holland) 563.  

\bibitem{Felder} G. Felder, in the Proceedings of the International 
Congress of Mathematicians, Z{\"u}rich, 1994 (Birkh{\"a}user, Basel, 1994) p.1247.     
\bibitem{FV1} G. Felder and A. Varchenko, Commun. Math. Phys. {\bf 181}(1996) 741 . 
\bibitem{FV2} G. Felder and A. Varchenko, Nucl. Phys. B {\bf 480} (1996) 485. 

\bibitem{ABF} G.E. Andrews, R.J. Baxter and P.J. Forrester, 
J. Stat. Phys. {\bf 35} (1984) 193. 

\bibitem{BBB} O. Babelon, D. Bernard, E. Billey, 
Phys. Lett. B {\bf 375} (1996) 89.  
\bibitem{Jimbo} M. Jimbo, H. Konno, S. Odake, J. Shiraishi, 
Transformation Groups {\bf 4}, 303 (1999).   




\bibitem{DFM} T. Deguchi, K. Fabricius and B.M. McCoy, J. Stat. 
Phys. {\bf 102} (2001) 701.   
\bibitem{FM} K. Fabricius and B.M. McCoy, 
J. Stat. Phys. {\bf 103}(2001) 647; 
J. Stat. Phys. {\bf 104}(2001) 575; cond-mat/0108057. 
%

\bibitem{Pearce} P.A.  Pearce and K.A. Seaton, Phys. Rev. Lett. {\bf 60} (1988) 1347. 
\bibitem{Kuniba} A. Kuniba and T. Yajima, J. Stat. Phys. {\bf 52}, 829 (1987). 
\bibitem{Akutsu} Y. Akutsu, T. Deguchi and M. Wadati, J. Phys. Soc. Jpn. {\bf 57}, 1173 (1988). 



\bibitem{XXX} T. Deguchi, J. Phys. A: Math. Gen. {\bf 34} (2001) 9755.   

\bibitem{Klaus} K. Fabricius, a private communication. 


\bibitem{A-IRF} M. Jimbo, T. Miwa and 
M. Okado, Nucl. Phys. B {\bf 300} (1988) 74. 
\bibitem{ABCD} M. Jimbo, T. Miwa and M. Okado, Commun. Math. Phys. 
{\bf 116} (1988) 353.  

\bibitem{Belavin} A.A. Belavin, Nucl. Phys. B {\bf 180} (1981) 189. 

\bibitem{Korff} C. Korff and B.M. McCoy, hep-th/0104120. 

\bibitem{Takebe} T. Takebe, J. Phys. A: Math. Gen. {\bf 28}(1995) 6675; 
J. Phys. A: Math. Gen. {\bf 29} (1996) 1563.   

\bibitem{Fuchs} J. Fuchs and C. Schweigert, 
{\it Symmetries, Lie Algebras and Representations}, 
(Cambridge Univ. Press, 1997). 
 

\end{thebibliography}
  \end{document}